\documentclass[10pt, letterpaper]{article}
\usepackage{graphics} 
\usepackage{epsfig} 
\usepackage{mathptmx} 
\usepackage{amssymb,amsthm}  
\usepackage{cite}
\usepackage{scalerel,stackengine}
\usepackage[tight,footnotesize]{subfigure}
\usepackage{mathtools}
\usepackage{balance}
\usepackage{color}
\usepackage[nolist,nohyperlinks]{acronym} 
\usepackage[tight,footnotesize]{subfigure}

\begin{document}
%
\title{
Detecting integrity attacks in IoT-based \\Cyber Physical Systems:\\ a case study on Hydra testbed
}
\date{}


\author{Federica Battisti, Giuseppe Bernieri,\\ Marco Carli, Michela Lopardo, and Federica Pascucci\\
Department of Engineering\\
Universit\`a degli Studi Roma Tre\\
Roma, Italy\\
 }

\maketitle

\begin{acronym}[] 
\acro{CPS}{Cyber-Physical System}
\acro{HMI}{Human Machine Interface}
\end{acronym}

\begin{abstract}
The Internet of Things paradigm improves the classical information sharing scheme. However, it has increased the need for granting the security of the connected systems. In the industrial field, the problem becomes more complex due to the need of protecting a large attack surface while granting the availability of the system and the real time response to the presence of threats. In this contribution, we deal with the injection of tampered data into the communication channel to affect the physical system. The proposed approach relies on designing a secure control system by coding the output matrices according to a secret pattern. This pattern is created by using the Fibonacci \textit{p-sequences}, numeric sequence depending on a key. The proposed method is validated on the Hydra testbed, emulating the industrial control network of a water distribution system.
\end{abstract}


%

\section{Introduction}

In the last decades, we assisted to the spread of a new trend aimed at connecting as many devices as possible. This trend led to what is commonly referred to as Industry 4.0, the fourth industrial revolution. 
The core innovation of this revolution is the possibility for the \ac{CPS} of exploiting Internet to extend the communication range beyond the closed industrial communication networks, thus leading to the birth of the Industrial Internet of Things (IIoT).
IIoT requires the deployment of sensors, actuators, and communication devices in the physical infrastructure for allowing the remote monitoring and control of the whole system as well as of its components.

The distributed nature of IIoT, together with the need of specific communication paradigms, and the adoption of Internet, may impact the reliability, the robustness, and the security of \ac{CPS}~\cite{Sajid_IEEE_2016}. 
In more details, in IoT \ac{CPS} context, the availability of the system is a key factor. In fact, since CPSs deal with dynamic systems, it is important that the service or system is always available and that the information is shared within time delays that can be very short in real time applications~\cite{Zeng_ISIE_2011}. Neverthless, timing and information reliability is of paramount importance: the detection of data modification (due to transmission errors or to malicious alteration) should rise an alert as soon as possible for a prompt reaction/mitigation. If data modification is not timely detected, it may result in severe disruption of the system or even in its complete damage.
Badly secured IIoT structures and services may be used as entry points for network attacks and expose both data and systems to threats~\cite{Rose_ISOC_2015, Stolpe_SIGKDD_2016, Rubio-Hernan_JIS_2017}.  

In this contribution, we aim at the design of a secure control system able to identify the injection of tampered data (e.g., the deception attack) in the communication channel. It useful to underline that the  integrity attack is extremely dangerous since it might be unnoticed till the unavailability of the physical system happens. 
We propose to code the physical output of the system through permutation matrices whose scheme varies based on a secret sequence. In more details, we extend the works in \cite{Miao_TCNS_2016,wimob} to non linear systems, by introducing the following innovations:
\begin{itemize}
\item the coding matrices are based on permutations obtained by rotation and flipping that modify the order of the elements in the output vector and their sign. To the best of our knowledge, in literature only a single rotation is used in~\cite{Miao_TCNS_2016}, and a small subset of signed permutation matrices in ~\cite{wimob}. In the proposed system, the flipping operation is adopted for increasing the number of possible matrices that can be used thus gaining in security; 
\item the encoding procedure satisfies the real time constraint and avoids quantization errors;
\item the computational complexity is highly reduced since the coding matrices can be precomputed off-line;
\item the security level of the system is increased by updating the coding matrix according to a rule based on key-dependent sequences, the Fibonacci \emph{p-sequences}, exploiting the communication protocol to avoid synchronization problems.
\end{itemize}

\section{Related works}
\label{sec:related}
The complex structure of a connected CPS is exposed to several attacks in different points, thus resulting in a large attack surface\cite{Sadeghi_DAC_2015}. It is possible to identify three layers as potential goals of an attacker: human, network, software, and hardware layer.  

Several studies have been performed for assessing the security or mitigating the effects of an attack in a IoT-based CPS. In particular, secure control theory is used to estimate the impact of cyber threats on the physical plant\cite{Cazorla_SJ_2016}. 
Given the complexity of the problem, the methods proposed in literature are usually dedicated to counteract attacks that can be roughly classified in two groups: DoS and deception-based attacks.
The impact of a DoS attack, even if limited to a subset of the network, may have a disruptive effect on the whole system \cite{Srikantha_ISGT_2015}. In \cite{Cardenas_RCS_2008}, the availability of the system or service when their functionalities are interrupted (i.e., by limiting the exchange of information between sensors and control system) is considered.
 In \cite{TEIXEIRA_Automatica_2015}, the attacker goal is to limit the availability of a subset of controls and sensors. To avoid the detection of his/her malicious behavior, the attacker mimics poor network conditions (i.e., by randomly dropping packets). Different mitigation methods are presented in \cite{Robinson_ICIEA_2012}.\\
In the deception-based attacks, the adversary, after having gained access to the CPS, injects false or tampered information towards or from sensors or controllers (i.e., the value of a measurement or the sensor identification label). An effective attack is designed to remain unnoticed to the detection system until a severe fault occurs.  In \cite{Kwon_ACC_2013}, the cases in which a stealthy deception attack may be performed without being detected are addressed.
In \cite{kim_TSG_2011}, the authors show that resiliency to malicious data injection may be obtained if a subset of measurements is immune to the attacks. 
In \cite{Mo_SCS_2010}, a false data injection attack model is presented as a constrained control problem and the theoretical analysis of the conditions  under which the attacker could successfully destabilize the system are shown.
An extended review of the security aspects is in \cite{Zhang_TIFS_2017}. 
In \cite{Miao_TCNS_2016} a smart attack on linear time-invariant systems is addressed. The data injection is performed in such a way that the state estimation error increases without being detected till the moment when the presence is fatal for the system itself. The authors propose to code the sensor measurements exploiting a Givens rotation matrix for securing the system output. A similar approach is adopted in \cite{wimob}. The key idea is to use a subset of the signed permutation matrices to perform rotation and flipping of the output vector space. In this way, the signal injected by the attacker, once decoded, introduces a large residual error, so it does not show the same statistical property of the healthy signal. 
The same problem is addressed in \cite{Ganesan_WSN_2003}, where a solution based on the encryption of shared information to protect data integrity (and confidentiality) is proposed. 
In our contribution we deal with the issues highlighted in \cite{Miao_TCNS_2016, wimob, Ganesan_WSN_2003}. In particular, we change the output coding scheme at each transmission time by selecting the coding matrix out of a predefined set, i.e., the set of the signed permutation matrices. This coding scheme avoids  quantization errors and reduces the computational complexity, since it results in scrambling the elements of the observation vector. 

\section{System and Adversary Modelling}
\label{sec:Definition}
\subsection{Industrial \ac{CPS}}
\begin{figure}[!t]
\centering\includegraphics[width=0.65\linewidth]{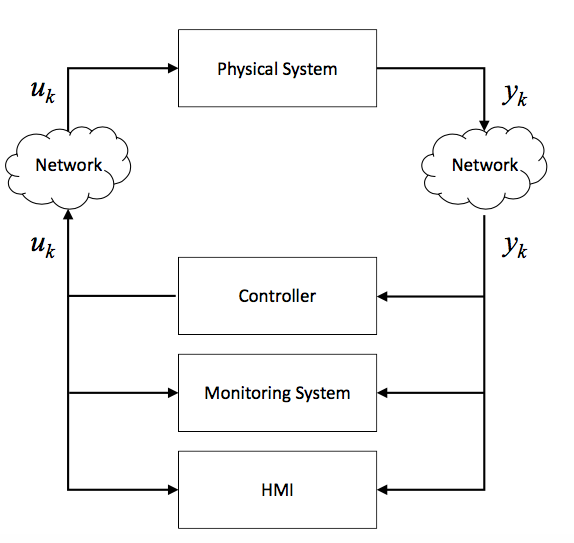}
\caption{The IoT based CPS and the detection scheme: the physical system communicates with the controller, the monitoring system, and the HMI by a network.}
\label{fig:propSystem}
\end{figure}

In this work the industrial \ac{CPS} depicted in Fig.~\ref{fig:propSystem} is considered.
It is composed by the \textit{physical system} (i.e., the actuators, the plant, and the sensors) and the \textit{monitoring system}, the \textit{controller},  and a \textit{\ac{HMI}}.
According to the IIoT paradigm, the sensors and the actuators of the physical system forward the collected data to the monitoring system, the controller, and the \ac{HMI} through the network at each transmission time $k$ using a protocol that identifies the sequence number of the packet in the data stream (e.g., Modbus/TCP).

The \textit{physical system} can be described by
a nonlinear uncertain system: thus, the discrete time model is given~by
\begin{equation}
\begin{array}{rcl}
x_{k} &=& f(x_{k-1},u_{k-1})+w_k\\
y_{k} &=& g(x_k)+v_k
\end{array}
\label{eq_sys}
\end{equation}
where $x_{k}\in \mathbb{R}^{n_x}$ is the state of the system, $u_{k}\in \mathbb{R}^{n_u}$ is the input, $y_{k}\in \mathbb{R}^{n_y}$ is the output, $f(\cdot)$ is the state transition map, $g(\cdot)$ is the observation map, and $w_k\in \mathbb{R}^q$, $v_k\in \mathbb{R}^l$ are the process and measurement noises, respectively. The input $u_k$ is known, while $w_k$ and $v_k$ are Gaussian white noise with known constant covariance matrices (i.e., ${w_k\sim \mathcal N (0,Q)}$ and $v_k\sim \mathcal N (0,R)$, respectively).

The \textit{monitoring system} is able to detect both faults and attacks. To this aim, it is implemented as a fault detection system~\cite{UnArtNostro} and it is composed by a state observer and a detector. The state observer is able to replicate the behavior of the plant (i.e., the estimate of the state $\hat{x}_{k}$), knowing the input from the controller and the output from the sensors at each transmission time $k$. It is implemented by means of an Extended Kalman Filter (EKF),  which estimates the state according to a prediction/correction scheme.
In the prediction, an a priori estimate $\hat{x}^{(-)}_{k}$ and its covariance  $P_k^{(-)}$ is computed as:
\begin{equation}
\begin{array}{rcl}
\hat{x}^{(-)}_{k} &= &f(\hat{x}_{k-1},u_{k-1})  \\
P_k^{(-)} &=& F P^{(+)}_{k-1} F^{T} + Q 
\end{array}
\end{equation}
where $F$ is the Jacobian of the state transition map $f(\cdot)$. The update of the estimate and its covariance are obtained as
\begin{equation}
\begin{array}{lcl}
K_{k} &= &(P_k^{(-)} G^{T})(G P_k^{(-)} G^{T} + R)^{-1}  \\
\hat{x}^{(+)}_{k} &= &\hat{x}^{(-)}_{k} + K_{k} ({y}_{k} - g(\hat{x}^{(-)}_{k}) \\
P_k^{(+)} &=& (I - K_{k} G) P_k^{(-)} 
\end{array}
\end{equation}
where $K_{k}$ represents the Kalman gain, $G$ is the Jacobian of the observation map $g(\cdot)$ , and $I$ is the identity matrix. Due to the nonlinearity of the system, a validation gate based on Mahalanobis distance and $\chi$-square test is implemented to exclude the outliers and avoid non-convergence. The update is performed only when the following inequality holds
\begin{equation}
[{y}_{k} - g(\hat{x}^{(-)}_{k})]^T(G P_k^{(-)} G^{T} + R)[{y}_{k} - g(\hat{x}^{(-)}_{k})]\leq \chi^2.  
\end{equation}
The detector evaluates the residual
\begin{equation}
r_{k}={y}_{k} - g(\hat{x}^{(+)}_{k}) 
\end{equation}
by comparing it with a threshold $\beta\in \mathbb{R}^{n_y}$ computed during a fault/attack free operating condition during the time interval $[0,\dots,T)$, so that
\begin{equation}
\beta_{i}=\max_{k=0,\dots, T}r_{k,i}  \,\,\,\,\,\, \forall i=1,\dots,n_y. 
\end{equation}
Finally, the following decision rule $\mathcal{R}_k$ is applied
\begin{equation}
\mathcal{R}_k=
\begin{cases}
\mathcal{H}_0 \quad \text{if } r_k\leq \beta \\
\mathcal{H}_1 \quad \text{if } r_k> \beta \\
\end{cases}
\end{equation}
where $\mathcal{H}_0$ is the \textit{healthy} and $\mathcal{H}_1$ is the \textit{under attack} hypothesis, respectively. When $\mathcal{H}_1$ is accepted, the monitoring system triggers an alarm and forwards the information to the~HMI.

The \textit{controller} is devoted to regulate the desired output implementing a feedback control law and it is represented by a Programmable Logic Controller (PLC). The HMI is represented by a Supervisory Control And Data Acquisition system.

\subsection{Adversary model}
In this work, the adversary is assumed to know the network topologies and the resources connected (i.e., the controller, the monitoring system, the sensors, and the actuators). The adversary objective is to reduce the availability of the resources by compromising data integrity. To this end, the target of the attacker is the communication channel between the sensors and the controller: the adversary is able manipulate the controller by injecting false data into this channel. Therefore, the adversary is assumed to be able to corrupt the communication channel between the concentrator and the state observer bypassing the attack detection tool (e.g., a conventional intrusion detection system). According to \cite{TEIXEIRA_Automatica_2015}, the \textit{disruption resources} of the adversary encompass both the plant and the monitoring systems, the \textit{disclosure resources} exploited during the attack are represented by the data in the communication channels, and the \textit{model knowledge} is not required.

The challenge of the attack is to remain stealthy with respect to the monitoring system. In \cite{Mo_SCS_2010} and \cite{Kwon_ACC_2013}, the conditions under which a stealth attack can be successfully set up are presented, however they consider only linear time invariant systems.
For stable non-linear systems, a stealth attack can be set up to get insights on the vulnerabilities of the network. For example, the adversary can set up a replay attack exploiting steady state output to test if the man in the middle attack is successful: at steady state, indeed, the output does not change and the monitoring system can be easily misled.

\section{Detection strategy}
\label{sec:proposed}
As previously mentioned, the proposed approach protects the communication channel between sensors and controller/monitoring system by
coding the system output, according to a secret and predefined pattern.
More specifically, the coding scheme is obtained by modifying the order of the elements in the output vector and eventually their sign. It is implemented by multiplying the output of the system, collected by the concentrator, with a \textit{signed permutation matrix} having only one non-zero entry (either $1$ or $-1$) in each row and column. The signed permutation matrices form a group with integer inverse, thus the encoding procedure does not introduce quantization errors.

The coding matrix is modified at each transmission time according to a shared key, that depends on the Fibonacci \textit{p-numbers} and on the packet number. The set of all the signed permutation matrices $\mathcal S_\Pi$ of a vector $y_k\in \mathbb R^m$ is generated and sorted: at each transmission time, the shared key is used to select the coding matrix from the sorted set.

In more details, security is given by:
\begin{itemize}
\item the seed of the sequence used for selecting the coding matrices;
\item the sorting of the set $\mathcal S_\Pi$;
\item the output coding (i.e., scrambling).
\end{itemize}

The coding matrices are continuously updated according to a rule based on the Fibonacci \emph{p-sequences}.
A Fibonacci \emph{p-sequence} $F_{p}(n)$ is defined by the following recursive formula:
\begin{equation}
F_{p}(n)=\left\{
         \begin{array}{ll}
           0, & n <0; \\
           1, & n=0; \\
           F_{p}(n-1)+F_{p}(n-p-1), & otherwise.
         \end{array}
       \right.
\end{equation}

Since the number of feasible rotations and flipping performed to obtain the coded output is limited to $n_y$, there is the need for mapping the selected Fibonacci \emph{p-sequence} to the interval $[1,\dots,n_y]$. In order to do this, the modulo operation with base $n_y$ is performed.

It should be noticed that, as demonstrated in~\cite{Wall_AMM_1960}, the sequence $F_{p}$(mod ${n_y}$) forms a periodic series, that is, it repeats by returning to its starting values. This could be a security issue since an eavesdropping could reveal the adopted secret sequence. To cope with this situation, in the envisaged system the sequences are periodically changed, although in this contribution the attacker is supposed to set up a blind replay attack. The period depends on the dimension of the observation space ${n_y}$ and the $p$ and can be easily computed~\cite{Wall_AMM_1960}.

Overall, the use of these sequences grants an increased security to the system thanks to two elements:
\begin{itemize}
\item it avoids the problem of synchronization in case of packet loss; in fact, the selected \textit{n} depends on the sequence number of the packet in the data stream;
\item the order of the matrices used for coding the output signal depends on the selected Fibonacci \emph{p-sequence}; by changing the \textit{p-value}, the order can be modified without increasing the computational complexity of the system.
\end{itemize}

The proposed detection strategy improves the state of the art under several perspectives.
It adopts the same approach as proposed in~\cite{Miao_TCNS_2016}. Instead of being encrypted~\cite{Ganesan_WSN_2003}, the outputs are coded. In fact, the encryption of each message requires increased computational complexity that may be not affordable in a real time and low energy consumption distributed system~\cite{Goldreich_FCV_2004}, especially when considering legacy systems. Moreover, with respect to~\cite{Miao_TCNS_2016}, we update the coding matrix at each transmission time thus reducing the probability of disclosure of the matrix. In~\cite{Miao_TCNS_2016} the coding matrix is updated periodically since its computation is hard. Furthermore, this approach is applied to nonlinear systems, that have not been yet considered in the literature.

\section{Case Study on Hydra testbed} 

The testbed Hydra~\cite{Ber2016} has been used to validate the proposed monitoring system. The Hydra testbed emulates a water distribution system that combines gravity and pumps to move the fluid inside the system. The physical structure of the testbed has been designed using a low-cost approach, however, it is interfaced to the control system by an industrial PLC over a Modbus/TCP network. In the following the testbed is described and a simple replay attack is performed.

\subsection{Hydra testbed}
\begin{figure}[!t]
\centering
\includegraphics[width=0.7\linewidth]{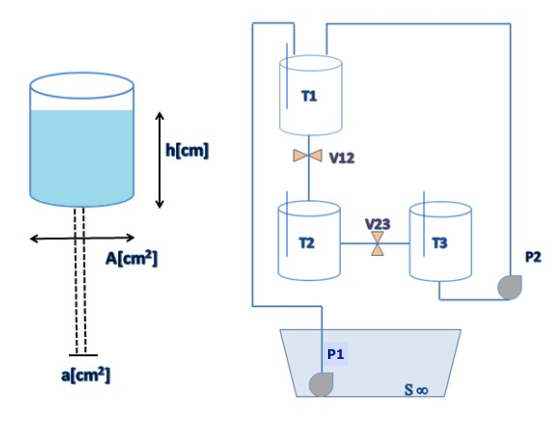}
\caption{The physical system of the Hydra testbed: the left-hand figure shows the single tank system, the right-hand one the whole system.}
\label{fig:testbed}
\end{figure}
The \textit{physical system} of the testbed is composed by $3$ tanks and a reservoir (see Fig.~\ref{fig:testbed}). Tanks $1$ and $2$ are connected by a serial pipeline: the fluid cascades due to gravity and the flow is regulated by the proportional valve $v_{1,2}$. Tank $2$ and $3$ are connected in a parallel configuration: the fluid moves due to Stevin's Law (communicant vessels) and the flow is regulated by the proportional valve $v_{2,3}$. Each tank is equipped with two redundant level sensors: the first one is represented by a pressure sensor, the second one by a sonar sensor.
The whole system is fed by a reservoir: the centrifugal pump $P_1$ provides water to Tank $1$, while the centrifugal pump $P_2$ links Tank $3$ and $1$. 

The \textit{SCADA} is implemented using Mango Automation that provides also a HMI. The \textit{controller} is developed using a Modicon M340 PLCs by Schneider Electric programmed in Ladder Logic using Unity Pro XL v7.0. It collects real-time data from the water level sensors and controls the actuators and executes the low-level control (e.g., it performs operator or SCADA commands, or the automatic maximum level control).
The proposed \textit{monitoring system} is implemented on a Galileo board. The continuous time state transition model is represented by the following equations:
\begin{equation}
\begin{array}{ccl}
A\dot{x_1} &=& P_1 + P_2 - Q_{1,2}\\
A\dot{x_2} &=&Q_{1,2}  - Q_{2,3} - Q_{2,3,h} + Q_{3,2,h}\\
A\dot{x_3} &=&  Q_{2,3} + Q_{2,3,h} - Q_{3,2,h} - P_2
\end{array}
\end{equation}
\noindent where:
\begin{small}
\begin{equation}
\begin{array}{ccl}
Q_{1,2} &=& av_{1,2}\sqrt{2gx_1}\\
Q_{2,3}&=&av_{2,3}\delta_{-1}(x_2 - h_{con})\delta_{-1}(x_3 - h_{con})\cdot\\
&&\text{sign} (x_2 - x_3)\sqrt{2g|x_2-x_3|}\\
Q_{2,3,h}&=&av_{2,3}\delta_{-1}(x_2 - x_{con})\delta_{-1}(x_{con} - x_3)\sqrt{2g(x_2-h_{con})}\\
Q_{3,2,h}&=& av_{2,3}\delta_{-1}(x_3 - x_{con})\delta_{-1}(x_{con} - x_2)\sqrt{2g(x_3-h_{con})}\\
P_2&=&k_2a\sqrt{2gx_3}\\
P_1&=& k_1
\nonumber
\end{array}
\end{equation}
\end{small}
$Q_{i,j}$ is the flow through tanks $i$ and $j$, $\delta_{-1}(\cdot)$ is the step signal, $g$ the gravitational acceleration, $h_{con}$ is the height of the connection between tanks $2$ and $3$, $k_i$ the gain of the $i$-th pump, $A$ and $a$ the area of the cross-section of the tanks and the pipes, respectively.
The discrete-time version of this model is used as state transition map in the prediction step of the EKF. The observation map is given by:
\begin{equation}
y_k=g(x_k) = I x_k.
\end{equation}
\begin{figure}[!t]
\centering
\includegraphics[width=0.5\linewidth]{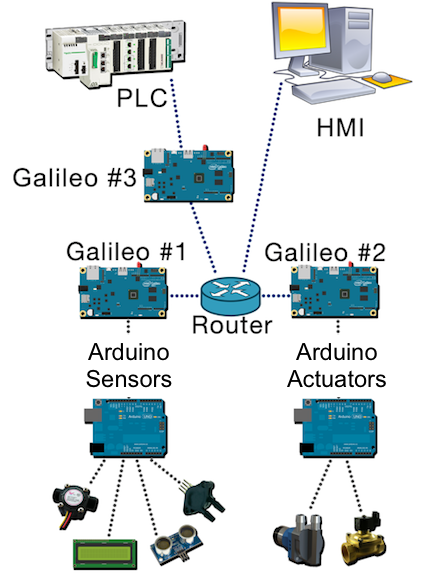}
\caption{The communication architecture of the Hydra testbed.}
\label{fig:testbed2}
\end{figure}

The network architecture of the Hydra testbed is shown in Fig.~\ref{fig:testbed2}: two Arduino boards and two Galileo boards control all the sensors and actuators. Specifically, one Arduino board is devoted to control the proportional valves and the centrifugal pumps that represent the actuators of the water distribution systems. The second Arduino board is used to interface the level sensors.
The Galileo boards are used to interface the Arduino ones on a ModBus/TCP network. 
The first one, Galileo \#1, collects data from the Arduino board devoted to interface the sensor and dispatch the measurements to the network by encapsulating the measurements in Modbus/TCP packets after applying the proposed coding scheme. The second one, Galileo \#2 is connected to the Arduino devoted to control the motors. It collects the input from the controller and forwards it to the corresponding Arduino board. 
Finally, another Galileo, Galileo \#3, is connected to the PLC: it runs the monitoring system and is able to decode the data from the sensors. All the modules are connected by a local network by means of an Ethernet router.

\subsection{Validation test}
\begin{figure*}[!t]
\centering 
\subfigure[Output without encoding.]{\label{fig:notcode} \includegraphics[width=0.45\textwidth, angle =0]{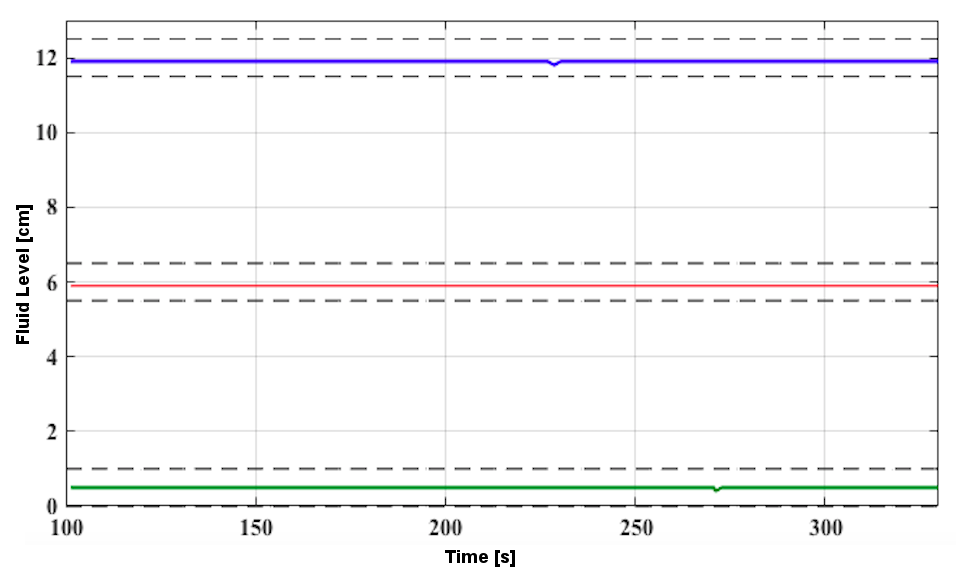}}
\hspace{0.7truecm}
\subfigure[Coded Output.]{\label{fig:output} \includegraphics[width=0.46\textwidth, angle =0]{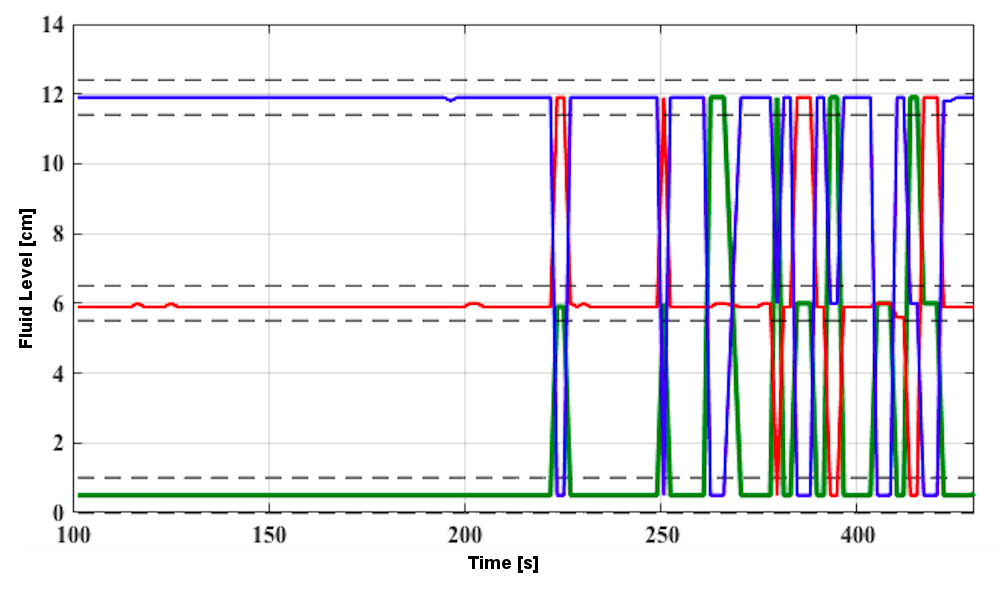}}\\
\label{out}
\end{figure*}
To prove the effectiveness of the proposed approach, an integrity attack has been set up. By means of a MITM attack, the communication link between the Galileo~\#1 and \#3 has been corrupted by injecting false data. 
We wrote code that exploits a vulnerability of the network to implement the attack and 
Wireshark had been exploited to collect network packets.

The attack starts using ARP cache poisoning, in this way the links between IP addresses and MAC addresses in the ARP table of the hosts (i.e., the Galileo \#1 and the Galileo~\#3) are corrupted. Consequently, the data stream from the Galileo~\#1 to the Galileo~\#3 is redirected to the malicious agent. The replay attack starts when the system reaches the steady state. This status of the system can be easily identified by eavesdropping the actuator controls, since these signals do not change when the system is at steady state.

During the attack, the adversary forwards to the Galileo~\#3 (and the PLC) wrong information about the level of fluid in the tanks. Specifically, the malicious agent replicates at each instant the levels recorded when the attack started. The result of the replay attack without coding the output is reported in Fig.~\ref{fig:notcode}: in this case the attack cannot be perceived by the monitoring system, since the attacked measurements look like the expected ones and are inside the tolerance introduced by the threshold. In this case, the monitoring system does not provide any alarm to the SCADA system and the attack can escalate. 

On the contrary, by applying the proposed coding method, the monitoring system is able to timely identify the attack.
The output analysis is shown in Fig~\ref{fig:output}: as it can be seen, the monitoring system clearly identifies the anomalies. The output vector changes at each transmission time since the attacker is not able to reproduce the correct scrambling sequence. As a result, all the thresholds of the residuals related to the state of each tank are violated and the detector triggers the alarm.

\section{Conclusion}
IoT based CPSs represent a revolution in many sectors: from industrial plants or energy generation systems, to distributed health systems. Smart services, cost production reduction, and quality assessment are only few of the possible advantages that results from this industrial revolution. However, due to the interdisciplinary nature of these systems, they are prone to security flaws. In this context, cyber attacks may result in physical damage of the system up to creating threats to the human life.
In this contribution, a method for securing IoT based CPSs through the timely detection of deception attacks is presented. The proposed approach is based on coding the output of the system by using permutation matrices selected from a set. This set is obtained through operations of scrambling and flipping. The selection of the permutation matrix is based on Fibonacci \textit{p-sequences}. The proposed detection strategy can cope with all issues related to the computational complexity of the coding step thus assuring the compliance with the time delay constraints typical of CPSs. Furthermore, quantization errors, that may have a nonlinear behavior and can compromise the convergence of the residual estimator, are avoided. The approach has been validated on an testbed that emulates industrial control systems. Future work will be devoted to apply more complex coding scheme to identify attacks that can be mislead by the monitoring system.

%
%
%



%

\end{document}